# ARE CRYPTOCURRENCY TRADERS PIONEERS OR JUST RISK-SEEKERS? EVIDENCE FROM BROKERAGE ACCOUNTS

Matthias Pelster*

University of Paderborn

Bastian Breitmayer*

Queensland University of Technology

Tim Hasso*†

Bond University

14 University Dr, Robina QLD 4226, Australia

thasso@bond.edu.au

**Abstract**

Are cryptocurrency traders driven by a desire to invest in a new asset class to diversify their portfolio or are they merely seeking to increase their levels of risk? To answer this question, we use individual-level brokerage data and study their behavior in stock trading around the time they engage in their first cryptocurrency trade. We find that when engaging in cryptocurrency trading investors simultaneously increase their risk-seeking behavior in stock trading as they increase their trading intensity and use of leverage. The increase in risk-seeking in stocks is particularly pronounced when volatility in cryptocurrency returns is low, suggesting that their overall behavior is driven by excitement-seeking.

Keywords: cryptocurrencies, bitcoin, investor, risk-seeking

JEL classifications: D1; G1

*All authors contributed equally, author order is the result of randomization

† Corresponding author

# 1. INTRODUCTION

Driven by increased participation and volatile markets, cryptocurrencies have received an increasing amount of attention both by academic and financial press in the past few years (Phillip, Chan, & Peiris, 2018; Urquhart, 2018). One of the major criticisms is that cryptocurrencies have neither fundamental value nor central bank association and that, therefore, trading cryptocurrencies cannot be seen as investing but instead as a form of speculation that is similar to gambling (Schiller, 2017). On the other hand, defenders have stated that cryptocurrencies are here to stay and are a valid new asset class, allowing investors to increase portfolio diversification (Brière, Oosterlinck, & Szafarz, 2015).

In this paper, we explore the motives of investors who trade cryptocurrencies. Is cryptocurrency trading driven by an intentional strategy to enter a new promising asset class and diversify, or is it merely part of an overall strategy of increasing risk-taking? To answer this question, we study the stock trading behavior of cryptocurrency traders around the time they execute their first cryptocurrency trade. The underlying notion is that if cryptocurrency traders are investing in cryptocurrencies as part of a risk-seeking strategy, then we should also see an increase in their stock trading intensity and risk-taking. Whereas if cryptocurrency traders are merely investing in a new promising asset-class, then their stock trading behavior should remain unchanged. We utilize a sample of 96,054 individual accounts from an online broker, which allows its clients to trade contracts for difference (CFD) on a variety of underlying instruments, including equities, and cryptocurrencies. We find that investors increase risk-taking on the stock market when they start trading cryptocurrencies.

The existing academic research in this area is increasing at a high rate. However, the majority of research focuses on market level information (Baek & Elbeck, 2015), which can be explained by the wide availability of market-level price information for cryptocurrencies.

Furthermore, the majority of the existing academic research on cryptocurrencies focuses on Bitcoin, whereas other cryptocurrencies have only recently been considered (Gkillas & Katsiampa, 2018). We extend prior research on market dynamics and instead focus on the individuals who buy and sell cryptocurrencies and their trading behavior. As such, we build on the work that has to date used search and brokerage data to understand who engages in cryptocurrency trading (Hasso, Pelster, & Breitmayer, 2019; Urquhart, 2018; Yelowitz & Wilson, 2015), and similarly to Baur and Dimpfl (2018) we contribute evidence in an area where economic models have previously been proposed but not empirically tested.

## 2. DATA

We use transactional-level brokerage data from an online broker that allows its customers to trade CFDs on a variety of instruments, including equities and cryptocurrencies[1] under a UK broker license. By being able to trade CFDs on cryptocurrencies with their trusted broker, investors can participate in cryptocurrency trading without having to worry about the risk of direct cryptocurrency investments, such as hacking or scams. CFD based trading is interesting from an investor behavior perspective as they can trade cryptocurrencies in a similar manner to how they already trade other asset classes. The sample period is from January 1, 2014, to December 31, 2017, and our initial dataset consists of 668,067 individual investors.

## 3. METHOD

We compare the trading activity of investors who start to trade cryptocurrencies to the trading activity of comparable investors who do not in a standard difference-in-differences setting. We restrict our analysis on the immediate trading activity of investors around the first time an investor starts to trade cryptocurrencies (treatment date) to mitigate other time-varying influences on investors' trading activity (20 days).

---

[1] The cryptocurrencies traded through the broker are Bitcoin, Bitcoin Cash, Dash, Ethereum, Ethereum Classic, Litecoin, Neo, Ripple, and Stellar.

We obtain our group of comparable investors from all investors who trade with the brokerage service and have not started to trade cryptocurrencies previous to the treatment date of the cryptocurrency trader. For the group of comparable investors, we obtain the trading activity before the treatment date and compare the trading activity to that of our cryptocurrency traders. We run a nearest-neighbor matching routine to match investors based on their previous trading intensity and performance, the treatment time, gender, age group, and trading characteristics. The trading characteristics include the average leverage the investors use, their average holding period, and average investment as a fraction of total portfolio value. We exclude investors for which we cannot find a match in our data from the analysis.

Our final dataset contains 96,056 individual investors. Summary statistics (untabulated) indicate that treated and comparable investors do not significantly differ after the matching. Using the matched sample, we estimate the following model

$Y_{it} = \beta_1 \cdot \text{Post treatment}_{it} + \beta_2 \cdot \text{Crypto trader}_i + \beta_3 \cdot \text{Post treatment}_{it} \cdot \text{Crypto trader}_i + \text{Controls}_{it} + e_{it},$

where Post treatment is a dummy variable that equals one after the treatment, regardless of whether investor i is the crypto trader or comparable trader, and zero otherwise. The variable captures all effects that are relevant for both types of investors after the treatment. Crypto trader is a dummy variable equal to one for crypto traders. It controls for all time-invariant differences between the two types of investors that are not accounted for by the matching routine. The interaction term between Post treatment and Crypto trader is the central term as its coefficient captures the change in risk-taking of treated investors that is related to starting to trade cryptos.

When estimating equation (1), we specify that the error term contains a fixed effect for investor, which is perfectly collinear with the dummy variable Crypto trader and controls for unobservable trader characteristics. We also include time fixed effects to control for aggregate time-trends. We control for investors' past profitability, their holding periods, the level of leverage, and the degree of diversification. Table 1 provides our variable definitions.

## 4. RESULTS

We present our primary results in Table 2. Models 1 to 3 consider the effects of cryptocurrency trading on the stock trading intensity, the level of leverage when trading stocks, and the average stock return. Models 4 to 6 consider the moderating effect of gender, whereas Models 7 to 9 consider the moderating effects of Age.

Our result show that, on average, cryptocurrency traders execute 16.8 additional stock trades within the first 10 days after initiating their cryptocurrency activities (Model 1). We also find that they increase their use of leverage by 13.4 percent (Model 2). Together, these effects appear to lead to decreased returns of cryptocurrency traders by 5 percent per month (Model 3). Female traders experience a somewhat diminished increase in leverage compared to their male counterparts (Model 5), but we find no gender differences when it comes to the number of trades or returns. We find evidence of a non-linear relationship between age and number of trades as the age group of 55 to 64 years old increase their number of trades the most (Model 7). In contrast, when it comes to leverage, we find that the age group of 35 to 44 years old increased their leverage the most, followed by 25 to 34 year old investors. We find no effects for age when it comes to returns.

To further explore the motivation of investors to engage in crypto trading, we investigate the stock trading activities of investors with respect to the volatility of the cryptocurrency they begin to trade. If investors are risk-seeking, the results documented in Table 2 should be stronger when crypto volatility is low. For each treatment time and cryptocurrency, we estimate the recent volatility. We then include this volatility in our model as an explanatory variable and interact it with the treatment effect. Table 3 presents the results. The interaction effect is negative for number of trades and leverage (Models 10 and 11). When investors' risk and excitement-seeking is not satisfied by high cryptocurrency volatility, they seek higher risk in their stock trading activities. This indicates that traders may be merely seeking risk when they engage in cryptocurrency trading.

## 5. DISCUSSION

Evidence in financial economics suggests that (some) investors treat trading as a fun and exciting activity (Gao & Lin, 2015). We contribute to this literature and show that cryptocurrency traders are motivated by risk-seeking behavior. Specifically, we find that, on average, investors who enter crypto markets simultaneously increase their risk-taking in stock trading, where they increase the number of trades and their leverage, resulting in lower returns. In addition, we find that the risk-seeking behavior in stocks is diminished, when cryptocurrency volatility is high, again suggesting that these traders are seeking risk and excitement.

We acknowledge that the traders in our sample may not be representative of the overall cryptocurrency market and CFD traders may differ from investors purchasing cryptocurrencies directly. However, the brokerage service self-reports itself as the market leader for CFD based cryptocurrency trading, and, allows us to obtain transaction-level data across both cryptocurrency and stock trading for the same individuals in the same trading environment.

Table 1. Variable definitions of trading variables

| Variable | Definition |
|---|---|
| *Trades* | Number of equity trades executed over a ten day period |
| *Leverage* | Average portfolio leverage over a ten day period where a value of one denotes no leverage |
| *Return* | Mean monthly return on equity trades |

Table 2. Difference in differences results

|  | (1) Trades | (2) Leverage | (3) Return | (4) Trades | (5) Leverage | (6) Return | (7) Trades | (8) Leverage | (9) Return |
|---|---|---|---|---|---|---|---|---|---|
| Treatment | 16.80*** | 0.72*** | -0.05*** | 16.77*** | 0.73*** | -0.04*** | 14.09*** | 0.28 | 0.00 |
|  | (23.74) | (11.04) | (-3.49) | (23.84) | (15.74) | (-3.37) | (5.40) | (1.38) | (0.06) |
| Treatment * Female |  |  |  | 0.61 | -0.28*** | -0.02 |  |  |  |
|  |  |  |  | (0.61) | (-4.31) | (-0.7) |  |  |  |
| Treatment * 18-24 |  |  |  |  |  |  | -2.15 | 0.29 | -0.08 |
|  |  |  |  |  |  |  | (-0.82) | (1.44) | (-1.14) |
| Treatment * 25-34 |  |  |  |  |  |  | 2.04 | 0.51** | -0.03 |
|  |  |  |  |  |  |  | (0.78) | (2.54) | (-0.49) |
| Treatment * 35-44 |  |  |  |  |  |  | 4.33* | 0.52*** | -0.03 |
|  |  |  |  |  |  |  | (1.65) | (2.61) | (-0.47) |
| Treatment * 45-54 |  |  |  |  |  |  | 6.83** | 0.25 | -0.02 |
|  |  |  |  |  |  |  | (2.56) | (-1.26) | (-0.3) |
| Treatment * 55-64 |  |  |  |  |  |  | 8.84*** | 0.14 | -0.06 |
|  |  |  |  |  |  |  | (2.83) | (-0.64) | (-0.83) |
| Trader-fixed effects | Yes | Yes | Yes | Yes | Yes | Yes | Yes | Yes | Yes |
| Time-fixed effects | Yes | Yes | Yes | Yes | Yes | Yes | Yes | Yes | Yes |
| $R^2$ | 0.24 | 0.14 | 0.04 | 0.24 | 0.14 | 0.04 | 0.24 | 0.14 | 0.04 |
| N | 96,056 | 96,056 | 96,056 | 96,056 | 96,056 | 96,056 | 96,056 | 96,056 | 96,056 |

The table presents OLS regressions. Column headings specify the dependent variable. Standard errors are double-clustered over investors and time. *t*-statistics are in parentheses.

*** p<0.01, ** p<0.05, * p<0.1

Table 3. Moderating effects of cryptocurrency volatility

|  | (10) Trades | (11) Leverage | (12) Return |
|---|---|---|---|
| Treatment | 19.77*** | 0.91*** | -0.06** |
|  | (21.59) | (9.98) | (-2.21) |
| Crypto volatility | 1.04 | 0.02 | 0.03 |
|  | (1.37) | (0.30) | (0.84) |
| Treatment * Crypto volatility | -4.59*** | -0.29*** | 0.00 |
|  | (-3.10) | (-2.81) | (0.14) |
|  |  |  |  |
| Trader-fixed effects | Yes | Yes | Yes |
| Time-fixed effects | Yes | Yes | Yes |
|  |  |  |  |
| $R^2$ | 0.26 | 0.74 | 0.12 |
| N | 96,056 | 96,056 | 96,056 |

The table presents OLS regressions. Column headings specify the dependent variable. Standard errors are double-clustered over investors and time. *t*-statistics are in parentheses.

*** p<0.01, ** p<0.05, * p<0.1